\documentstyle[12pt]{article}

\font\twlgot =eufm10 scaled \magstep1
\font\egtgot =eufm8
\font\sevgot =eufm7
\font\twlmsb =msbm10 scaled \magstep1
\font\egtmsb =msbm8
\font\sevmsb =msbm7

\newfam\gotfam
\def\pgot{\fam\gotfam\twlgot}
\textfont\gotfam\twlgot
\scriptfont\gotfam\egtgot
\scriptscriptfont\gotfam\sevgot
\def\got{\protect\pgot}
\newfam\msbfam
\textfont\msbfam\twlmsb
\scriptfont\msbfam\egtmsb
\scriptscriptfont\msbfam\sevmsb
\def\Bbb{\protect\pBbb}
\def\pBbb{\relax\ifmmode\expandafter\Bb\else\typeout{You cann't use
Bbb in text mode}\fi}
\def\Bb #1{{\fam\msbfam\relax#1}}

\newcommand{\gJ}{{\got J}}

\def\thebibliography#1{\section*{References}\list
  {[\arabic{enumi}]}{\settowidth\labelwidth{#1}\leftmargin\labelwidth
    \advance\leftmargin\labelsep
    \usecounter{enumi}}
    \def\newblock{\hskip .11em plus .33em minus .07em}
    \sloppy\clubpenalty4000\widowpenalty4000
    \sfcode`\.=1000\relax}

\def\op#1{\mathop{\fam0 #1}\limits}

\newcommand{\beq}{\begin{equation}}
\newcommand{\eeq}{\end{equation}}
\newcommand{\ben}{\begin{eqnarray}}
\newcommand{\een}{\end{eqnarray}}
\newcommand{\be}{\begin{eqnarray*}}
\newcommand{\ee}{\end{eqnarray*}}
\newcommand{\bea}{\begin{eqalph}}
\newcommand{\eea}{\end{eqalph}}
\newcommand{\cA}{{\cal A}}

\newcommand{\cL}{{\cal L}}

\newcommand{\cE}{{\cal E}}
\newcommand{\cH}{{\cal H}}

\newcommand{\cO}{{\cal O}}

\newcommand{\bL}{{\bf L}}

\newcommand{\al}{\alpha}

\newcommand{\dl}{\delta}
\newcommand{\la}{\lambda}

\newcommand{\f}{\phi}
\newcommand{\om}{\omega}
\newcommand{\Om}{\Omega}
\newcommand{\m}{\mu}

\newcommand{\g}{\gamma}
\newcommand{\G}{\Gamma}
\newcommand{\th}{\theta}

\newcommand{\si}{\sigma}

\newcommand{\w}{\wedge}

\newcommand{\wh}{\widehat}
\newcommand{\ol}{\overline}
\newcommand{\dr}{\partial}

\newcommand{\ot}{\otimes}
\newcommand{\ap}{\approx}

\newcommand{\bx}[1]{{\rm\fbox{$ #1$}}}

\newcounter{remark}
\newcounter{example}
\newcounter{theorem}
\newcounter{proposition}
\newcounter{lemma}
\newcounter{corollary}
\newcounter{definition}
\def\theremark{\arabic{remark}}

\def\thedefinition{\arabic{definition}}

\newenvironment{prop}{\refstepcounter{definition}
\bigskip\noindent{\bf Proposition \thedefinition.}\it}{\medskip}

\newcommand{\mar}[1]{}

\begin{document}
\hbox{}

{\parindent=0pt

{\large \bf Lagrangian and Hamiltonian dynamics of
submanifolds}
\bigskip

{\bf G.Giachetta}, {\bf L.Mangiarotti}$^1$, {\bf G.Sardanashvily}$^2$
\bigskip

\begin{small}

$^1$ Department of Mathematics and Informatics, University of
Camerino, 62032 Camerino (MC), Italy 

\medskip

$^2$ Department of Theoretical Physics, Moscow State University,
117234 Moscow, Russia

\bigskip

{\bf Abstract}
Submanifolds of a manifold are described as sections of a certain fiber
bundle that enables one to consider their Lagrangian and (polysymplectic)
Hamiltonian dynamics as that of a particular classical field theory. 
In particular, their Lagrangians and Hamiltonians must satisfy rather
restrictive Noether identities. For
instance, this is the case of relativistic mechanics and classical string
theory.

\end{small}
\bigskip

}

\section{Introduction}

As is well known, fiber bundles and jet manifolds of their sections
provide an adequate mathematical formulation of classical field theory.
In particular, field Lagrangians and their Euler--Lagrange operators are
algebraically described as elements of the variational bicomplex
\cite{tak2,ander,jmp01}. This description is extended to Lagrangian
theory of odd fields \cite{barn,cmp04,jmp05}. 
The Hamiltonian counterpart of first-order Lagrangian theory on fiber
bundles is covariant Hamiltonian formalism developed in
multisymplectic, polysymplectic, and Hamilton --  De Donder)
variants (see, e.g., \cite{book,jpa99,krup,ech,leon}).

Jets of sections of fiber bundles are particular jets of submanifolds.
Namely, a space of jets of submanifolds admits a cover by charts of jets
of sections \cite{book,kras,modu}. Three-velocities in relativistic
mechanics exemplify first order jets of submanifolds
\cite{sard98,book98}.  A problem is that differential forms on jets of
submanifolds do not constitute a variational bicomplex because horizontal
forms (e.g., Lagrangians) are not preserved under coordinate
transformations. 

We consider $n$-dimensional submanifolds of an $m$-dimensional smooth
real manifold $Z$, and associate to them sections of a trivial fiber
bundle $Z_Q=Q\times Z\to Q$, where $Q$ is some $n$-dimensional manifold.
Here, we restrict our consideration to first order jets of submanifolds,
and state their relation to jets of sections of the fiber bundle $Z_Q\to
Q$ (the formulas (\ref{s17}), (\ref{s31}), and Proposition 1). 
This
relation fails to be one-to-one correspondence. The ambiguity contains,
e.g., diffeomorphisms of $Q$. 
Then  Lagrangian and (polysymplectic)
Hamiltonian formalism on a fiber bundle $Z_Q\to Q$ is developed in a
standard way, but Lagrangians and Hamiltonians are required to be 
variationally invariant under the above mentioned diffeomorphisms of $Q$.
This invariance however leads to rather restrictive Noether identities
(\ref{s60}) and (\ref{s110}) which these Lagrangians and Hamiltonians
must satisfy, unless other fields are introduced.

In a different way, one can choose some subbundle of the fiber bundle
$Z_Q\to Q$ in order to avoid the above mentioned ambiguity between jets
of subbundles of $Z$ and jets of sections of $Z_Q\to Q$. Since such a
subbundle itself need not be a jet manifold of some fiber bundle, it is 
a nonholonomic constraint. For
instance, this is the case of relativistic mechanics, phrased in terms of
four-velocities.

\section{Jets of submanifolds}

Given an $m$-dimensional smooth real manifold $Z$, a
$k$-order jet of $n$-dimensional submanifolds of $Z$ at a point
$z\in Z$ is defined as the equivalence class $j^k_zS$ 
of 
$n$-dimensional imbedded submanifolds of $Z$ through $z$ which
are tangent to each other at $z$ with order $k\geq 0$.
Namely, two submanifolds
$i_S: S\hookrightarrow Z$,  $i_{S'}:
S'\hookrightarrow Z$ 
through a point $z\in Z$ belong to the same equivalence class $j^k_zS$ 
iff the images of the $k$-tangent morphisms
\be
T^ki_S: T^kS\hookrightarrow T^kZ, \qquad  T^ki_{S'}:
T^kS'\hookrightarrow T^kZ 
\ee 
coincide with each other.
The set $J^k_nZ=\op\bigcup_{z\in Z} j^k_zS$ of $k$-order jets is a
finite-dimensional real smooth manifold. 
One puts $J^0_nZ =Z$. If $k>0$, 
let $Y\to X$ be an $m$-dimensional
fiber bundle over an
$n$-dimensional base $X$ and $J^kY$ the $k$-order jet manifold of
sections of $Y\to X$ (or, shortly, the jet manifold of $Y\to X$).  Given
an imbedding
$\Phi:Y\to Z$, there is the natural injection   
\be
J^k\Phi: J^kY\to J^k_nZ, \qquad j^k_xs \mapsto [\Phi\circ
s]^k_{\Phi(s(x))}, 
\ee
where $s$ are sections of $Y\to X$. This injection defines a
chart on
$J^k_nZ$. These charts provide a manifold atlas of $J^k_nZ$.

Here, we restrict our consideration to first order jets of submanifolds.
There is obvious one-to-one correspondence 
\mar{s10}\beq
\zeta: j^1_zS \mapsto V_{j^1S}\subset T_zZ  \label{s10}
\eeq
between the jets $j^1_zS$ at a point $z\in Z$ and the $n$-dimensional
vector subspaces of the tangent space $T_zZ$ of $Z$ at $z$. It follows
that
$J^1_nZ$ is a fiber bundle 
\mar{s3}\beq
\rho:J^1_nZ\to Z \label{s3}
\eeq
in Grassmann manifolds. It possesses the following coordinate atlas.

 Let $\{(U;z^\m)\}$ be a coordinate atlas of $Z$. 
Though $J^0_nZ=Z$, let us provide
$J^0_mZ$ with the atlas obtained by replacing  every chart $(U,z^A)$ of
$Z$ with the 
\be
{m\choose n}=\frac{m!}{n!(m-n)!}
\ee
charts on $U$ which 
correspond to different partitions of $(z^A)$ in 
collections of $n$ and $m-n$ coordinates 
\mar{5.8}\beq
(U; x^a,y^i), \qquad a=1,\ldots,n,  \qquad i=1,\ldots,m-n.\label{5.8}
\eeq
The transition functions between the coordinate charts (\ref{5.8}) of
$J^0_nZ$ associated with a coordinate chart 
$(U,z^A)$ of $Z$ reduce to an exchange between 
coordinates $x^a$ and $y^i$.
Transition functions between arbitrary coordinate charts of the
manifold $J^0_nZ$  take the form 
\mar{5.26} \beq
x'^a = x'^a (x^b, y^k), \qquad y'^i = y'^i (x^b, y^k).
\label{5.26}
\eeq

Given an atlas of coordinate charts (\ref{5.8}) -- (\ref{5.26}) of the
manifold
$J^0_nZ$, the first order jet manifold $J^1_nZ$  is endowed with the
coordinate charts
\mar{5.31}\beq
(\rho^{-1}(U)=U\times\Bbb R^{(m-n)n}; x^a,y^i,y^i_a), \label{5.31}
\eeq
possessing the following transition functions. 
With respect to the coordinates (\ref{5.31}) on the jet manifold $J^1_nZ$
and the induced fiber coordinates $(\dot x^a, \dot y^i)$ on the
tangent bundle $TZ$, the above mentioned correspondence $\zeta$
(\ref{s10}) reads
\be
\zeta: (y^i_a) \mapsto \dot x^a(\dr_a +y^i_a(j^1_zS)\dr_i).
\ee
It implies the relations
\mar{s0,1}\ben
&&  y'^j_a= (\frac{\dr y'^j}{\dr y^k} y^k_b + \frac{\dr y'^j}{\dr x^b})
(\frac{\dr x^b}{\dr y'^i}y'^i_a + \frac{\dr x^b}{\dr x'^a}), \label{s0}\\
&& (\frac{\dr x^b}{\dr y'^i}y'^i_a + \frac{\dr x^b}{\dr x'^a})
(\frac{\dr x'^c}{\dr
y^k} y^k_b + \frac{\dr x'^c}{\dr x^b})=\dl^c_a,\label{s1}
\een
which jet coordinates $y^i_a$ must satisfy under coordinate
transformations (\ref{5.26}). Let consider a nondegenerate $n\times n$
matrix $M$ with the entries
\be
 M^c_b=(\frac{\dr x'^c}{\dr
y^k}y^k_b + \frac{\dr x'^c}{\dr x^b}).
\ee
Then the relations (\ref{s1}) lead to the equalities 
\be
(\frac{\dr x^b}{\dr y'^i} y'^i_a + \frac{\dr x^b}{\dr x'^a})=
(M^{-1})^b_a.
\ee
Hence, we obtain the 
transformation law of first order jet coordinates
\mar{s2}\beq
 y'^j_a=
( \frac{\dr y'^j}{\dr y^k} y^k_b+ \frac{\dr y'^j}{\dr x^b})
(M^{-1})^b_a.
\label{s2}
\eeq
For instance, these are the Lorentz transformation of three-velocities
in relativistic mechanics.  In particular, if coordinate
transition functions $x'^a$ (\ref{5.26}) are independent of
coordinates $y^k$, the transformation law (\ref{s2}) comes to the familiar
transformations of jets of sections.

A glance at the transformations (\ref{s2}) shows that, in contrast
with a fiber bundle of jets of sections, the fiber bundle
(\ref{s3}) is not affine.  In particular, one generalizes the notion of
a connection on fiber bundles and treat global sections of the jet
bundle (\ref{s3}) as preconnections
\cite{modu}. However, a global section of this bundle need not
exist (\cite{ste}, Theorem 27.18).

Given a coordinate chart (\ref{5.31}) of $J^1_nZ$, one can regard
$\rho^{-1}(U)\subset J^1_nZ$ as the first order jet manifold $J^1U$ of
sections of the fiber bundle
\mar{s11}\beq
U\ni (x^a,y^i)\to (x^a)\in U_X. \label{s11}
\eeq 
The graded differential algebra $\cO^*(\rho^{-1}(U))$ of exterior forms
on 
$\rho^{-1}(U)$ is generated by horizontal forms $dx^a$ and contact forms
$dy^i-y^i_adx^a$. Coordinate transformations (\ref{5.26}) and
(\ref{s2}) preserve the ideal of contact forms, but horizontal forms
are not transformed into horizontal forms, unless coordinate
transition functions $x'^a$ (\ref{5.26}) are independent of
coordinates $y^k$. Therefore, one can
develop first order Lagrangian formalism with a Lagrangian $L=\cL d^nx$
on a coordinate chart
$\rho^{-1}(U)$, but this Lagrangian fails to be globally defined on
$J^1_nZ$.

In order to overcome this difficulty, let us consider an above mentioned
product 
$Z_Q=Q\times Z$ of $Z$ and an
$n$-dimensional real smooth manifold $Q$. We have a trivial fiber bundle
\mar{s12}\beq
\pi: Z_Q=Q\times Z\to Q, \label{s12}
\eeq
whose trivialization throughout holds fixed. This fiber bundle is 
provided with an atlas of coordinate charts
\mar{s20}\beq
(U_Q\times U; q^\m,x^a,y^i), \label{s20}
\eeq
where $(U; x^a,y^i)$ are the above mentioned coordinate charts
$(\ref{5.8})$ of the manifold
$J^0_n Z$. The coordinate charts (\ref{s20}) possess 
transition functions 
\mar{s21}\beq
q'^\m=q^\m(q^\nu), \qquad x'^a = x'^a (x^b, y^k), \qquad y'^i = y'^i
(x^b, y^k). \label{s21}
\eeq
Let $J^1Z_Q$ be the first order jet manifold of the fiber bundle
(\ref{s12}). Since the trivialization (\ref{s12}) is fixed, it is a
vector bundle $\pi^1:J^1Z_Q\to Z_Q$ isomorphic to the tensor
product 
\mar{s30}\beq
J^1Z_Q= T^*Q\op\ot_{Q\times Z} TZ \label{s30}
\eeq
of the cotangent bundle $T^*Q$ of $Q$ and the
tangent bundle $TZ$ of $Z$ over $Z_Q$. 

Given a coordinate atlas
(\ref{s20}) - (\ref{s21}) of 
$Z_Q$, the jet manifold  $J^1Z_Q$ is endowed with the
coordinate charts
\mar{s14}\beq
 ((\pi^1)^{-1}(U_Q\times U)=U_Q\times U\times\Bbb R^{mn};
q^\m,x^a,y^i,x^a_\m, y^i_\m), \label{s14}
\eeq
possessing transition functions
\mar{s16}\beq
x'^a_\m=(\frac{\dr x'^a}{\dr
y^k}y^k_\nu + \frac{\dr x'^a}{\dr x^b}x^b_\nu )\frac{\dr q^\nu}{\dr
q'^\m}, \qquad y'^i_\m=(\frac{\dr
y'^i}{\dr y^k}y^k_\nu + \frac{\dr y'^i}{\dr x^b}x^b_\nu)\frac{\dr
q^\nu}{\dr q'^\m}. \label{s16}
\eeq
Relative to coordinates (\ref{s14}), the isomorphism (\ref{s30}) takes
the form
\be
(x^a_\m, y^i_\m) \to dq^\m\ot(x^a_\m \dr_a + y^i_\m \dr_i).
\ee

Obviously, a jet $(q^\m,x^a,y^i,x^a_\m, y^i_\m)$ of
sections of the fiber bundle (\ref{s12}) defines some jet of
$n$-dimensional subbundles of the manifold
$\{q\}\times Z$ through a point $(x^a,y^i)\in Z$ if an $m\times n$
matrix with the entries $x^a_\m, y^i_\m$ is of maximal rank $n$.
This property is
preserved under the coordinate transformations (\ref{s16}).  
An element of $J^1Z_Q$ is called regular if it possesses this property.
Regular elements constitute an open subbundle of the jet bundle
$J^1Z_Q\to Z_Q$.

Since regular elements of $J^1Z_Q$ characterize jets of submanifolds of
$Z$, one hopes to describe the dynamics of
submanifolds of a manifold $Z$ as that of sections of the fiber bundle
(\ref{s12}).  For this purpose, let us refine the relation
between elements of the jet manifolds $J^1_nZ$ and $J^1Z_Q$.

Let us consider the manifold product $Q\times J^1_nZ$. Of course, it is
a bundle over $Z_Q$.  Given a coordinate atlas
(\ref{s20}) - (\ref{s21}) of 
$Z_Q$, this product is endowed 
with the coordinate charts
\mar{s13}\beq
(U_Q\times \rho^{-1}(U)=U_Q\times U\times\Bbb R^{(m-n)n};
q^\m,x^a,y^i, y^i_a), \label{s13}
\eeq
possessing transition functions (\ref{s2}). 
Let us assign to an element $(q^\m,x^a,y^i, y^i_a)$ of the chart
(\ref{s13}) the elements $(q^\m,x^a,y^i,x^a_\m, y^i_\m)$ of the
chart (\ref{s14}) whose coordinates obey the relations
\mar{s17}\beq
\bx{y^i_a x^a_\m = y^i_\m.} \label{s17}
\eeq
These elements make up an $n^2$-dimensional vector space. The relations
(\ref{s17}) are maintained under  coordinate transformations (\ref{s21})
and the induced transformations of the charts (\ref{s14}) and (\ref{s13})
as follows:
\be
&& y'^i_a x'^a_\m = 
( \frac{\dr y'^i}{\dr y^k} y^k_c+
\frac{\dr y'^i}{\dr x^c}) (M^{-1})^c_a
(\frac{\dr x'^a}{\dr
y^k}y^k_\nu + \frac{\dr x'^a}{\dr x^b}x^b_\nu)\frac{\dr q^\nu}{\dr q'^\m}
= \\ 
&& \qquad ( \frac{\dr y'^i}{\dr y^k} y^k_c+
\frac{\dr y'^i}{\dr x^c}) (M^{-1})^c_a
(\frac{\dr x'^a}{\dr
y^k}y^k_b + \frac{\dr x'^a}{\dr x^b} )x^b_\nu\frac{\dr q^\nu}{\dr q'^\m}
=\\
&& \qquad (\frac{\dr y'^i}{\dr
y^k}y^k_b + \frac{\dr y'^i}{\dr x^b})x^b_\nu\frac{\dr q^\nu}{\dr q'^\m}=
(\frac{\dr y'^i}{\dr
y^k}y^k_\nu + \frac{\dr y'^i}{\dr x^b}x^b_\nu)\frac{\dr q^\nu}{\dr
q'^\m}= y'^i_\m.
\ee
Thus, one can associate
\mar{s25}\beq
\zeta': (q^\m,x^a,y^i, y^i_a) \mapsto \{(q^\m,x^a,y^i,x^a_\m, y^i_\m) \,
| \, y^i_a x^a_\m = y^i_\m\} \label{s25}
\eeq
to each element of the manifold $Q\times J^1_nZ$ an
$n^2$-dimensional vector space in the jet manifold $J^1Z_Q$. This
is a subspace of elements $x^a_\m dq^\m\ot(\dr_a + y^i_a\dr_i)$ of a
fiber of the tensor bundle (\ref{s30}) at a point
$(q^\m,x^a,y^i)$. This subspace always contains regular elements, e.g.,
whose coordinates $x^a_\m$ form a nondegenerate $n\times n$ matrix.

Conversely, given a regular element $j^1_zs$ of $J^1Z_Q$, there is a
coordinate chart (\ref{s14}) such that coordinates $x^a_\m$ of
$j^1_zs$  constitute a nondegenerate matrix, and $j^1_zs$ defines a unique
element of
$Q\times J^1_nZ$ by the relations
\mar{s31}\beq
\bx{y^i_a=y^i_\m(x^{-1})^\m_a.} \label{s31}
\eeq
For instance, this is the well-known relation between three- and
four-velocities in relativistic mechanics.

Thus, we have shown the following. Let $(q^\m,z^A)$ further be arbitrary
coordinates on the product
$Z_Q$ (\ref{s12}) and $(q^\m,z^A,z^A_\m)$ the corresponding coordinates
on the jet manifold $J^1Z_Q$. In these coordinates, an element of
$J^1Z_Q$ is regular if an $m\times n$ matrix with the entries $z^A_\m$ is
of maximal rank $n$.

\begin{prop} \label{s50} \mar{s50}
(i) Any jet of submanifolds through a point $z\in Z$ defines some (but
not unique) jet of sections of the fiber bundle $Z_Q$ (\ref{s12}) through
a point $q\times z$ for any $q\in Q$ in accordance with the relations
(\ref{s17}). 

(ii)  Any regular element of
$J^1Z_Q$ defines a unique element of the jet manifold $J^1_nZ$ by means
of the relations (\ref{s31}).  However, nonregular elements of $J^1Z_Q$
can correspond to different jets of submanifolds.

(iii) Two elements $(q^\m,z^A,z^A_\m)$ and $(q^\m,z^A,z'^A_\m)$ of
$J^1Z_Q$ correspond to the same jet of submanifolds if $z'^A_\m= M^\nu_\mu
z^A_\nu$, where $M$ is some matrix, e.g., it 
comes from a diffeomorphism of $Q$.
\end{prop}

Basing on this result, we can describe the
dynamics of $n$-dimensional submanifolds of a manifold $Z$ as that of
sections of the fiber bundle $Q\times Z\to Q$ for some $n$-dimensional 
manifold $Q$.

\section{Lagrangian dynamics of submanifolds}

Let $Z_Q$ be a fiber bundle (\ref{s12}) coordinated by $(q^\m, z^A)$ with
transition functions $q'^\m(q^\nu)$ and $z'^A(z^B)$. Then the first
order jet manifold $J^1Z_Q$ of this fiber bundle is provided with
coordinates $(q^\m, z^A, z^A_\m)$ possessing transition functions
\be
z'^A_\m=\frac{\dr z'^A}{\dr z^B} \frac{\dr q^\nu}{\dr q'^\m} z^B_\nu.
\ee
Let us recall the notation of contact forms $\th^A=dz^A-z^A_\m dq^\m$, 
operators of total derivatives
\be
d_\m=\dr_\m + z^A_\m\dr_A + z^A_{\m\nu}\dr_A^\nu,
\ee
the total differential $d_H (\f)= dq^\m\w d_\m(\f)$ acting on exterior
forms $\f$ on $J^1Z_Q$, and the horizontal projection $h_0(\th^A)= 0$.

A first order Lagrangian in Lagrangian formalism on a fiber bundle
$Z_Q\to Q$ is defined as a horizontal density
\mar{s40}\beq
L=\cL(z^A, z^A_\m) \om, \qquad \om=dq^1\w\cdots \w dq^n, \label{s40}
\eeq
on the jet manifold $J^1Z_Q$. The corresponding Euler--Lagrange operator
reads
\mar{s41}\beq
\dl L= \cE_A dz^A\w \om, \qquad \cE_A= \dr_A\cL - d_\m \dr_A^\m\cL.
\label{s41}
\eeq
It yields the Euler--Lagrange equations
\mar{s90}\beq
\cE_A= \dr_A\cL - d_\m \dr_A^\m\cL =0. \label{s90}
\eeq

Let $u=u^\m\dr_\m + u^A\dr_A$ be a vector field on $Z_Q$. Its jet
prolongation onto $J^1Z_Q$ reads
\mar{s52}\beq
u= u^\m\dr_\m + u^A\dr_A + (d_\m u^A -z^A_\nu d_\m u^\nu)\dr_A^\m.
\label{s52}
\eeq
It admits the vertical splitting
\mar{s53}\beq
u= u_H + u_V= u^\m d_\m +
[(u^A-u^\nu z^A_\nu)\dr_A + d_\m(u^A -z^A_\nu u^\nu)\dr_A^\m].
\label{s53}
\eeq
The Lie derivative $\bL_uL$ of a Lagrangian $L$ along a vector field 
$u$ obeys the first variational formula
\mar{s42}\beq
\bL_uL = u_V\rfloor \dl L + d_H(h_0(u\rfloor H_L))=((u^A-u^\m
z^A_\m) \cE_A + d_\m
\gJ^\m)\om,
\label{s42} 
\eeq
where  
\mar{s43}\beq
H_L=L +\dr_A^\m\cL\th^A\w \om_\m, 
\qquad \om_\m=\dr_\m\rfloor \om, \label{s43} 
\eeq
is the Poincar\'e--Cartan form and
\mar{s44}\beq
\gJ=\gJ^\m\om_\m=( \dr^\m_A\cL(u^A- u^\nu z^A_\nu) + u^\m\cL)\om_\m
\label{s44}
\eeq
is the Noether current. A vector field $u$ is called a variational
symmetry of a Lagrangian $L$ if the Lie derivative (\ref{s42}) is
$d_H$-exact, i.e., $\bL_u L=d_H\si$. In this case, there is the weak
conservation law $0\ap d_H(\gJ-\si)$
on the shell (\ref{s90}). 

One can show that a vector field
$u$ (\ref{s52}) is a variational symmetry only if it is projected onto
$Q$, i.e., its components $u^\m$ are functions on $Q$, and iff its
vertical part $u_V$ (\ref{s53}) is a variational symmetry. 
In a general setting, one deals with generalized vector fields
$u$ depending on parameter functions $\xi^r(q^\nu)$, their derivatives
$\dr_{\la_1\ldots\la_k}\xi^r$, and higher order jets 
$z^A_{\la_1\ldots\la_k}$ \cite{cmp04,jpa05}. A vertical variational
symmetry depending on parameters is called a gauge symmetry. Here we
restrict our consideration to gauge symmetries 
$u$ which are linear in parameters and their first derivatives, i.e., 
\mar{s55}\beq
u= u^A\dr_A + d_\m u_A\dr^\m_A, \qquad u^A= u^A_r\xi^r +
u^{A,\m}_r\dr_\m\xi^r, \label{s55}
\eeq
where $u^A$ are functions of $q^\m$, $z^B$ and the jets
$z^B_{\m_1\ldots\m_k}$ of bounded jet order $k<N$. By virtue of the
Noether second theorem
\cite{jmp05,jpa05,jmp05a}, 
$u$ (\ref{s55}) is a gauge
 symmetry of a Lagrangian $L$ (\ref{s40}) iff the variational derivatives
$\cE_A$ (\ref{s41}) of
$L$ obey the Noether identities
\mar{s56}\beq
(u^A_r-d_\m u^{A,\m}_r)\cE_A - u^{A,\m}_r d_\m\cE_A=0. \label{s56}
\eeq

For instance, let us consider an arbitrary  vector field
$u=u^\m(q^\nu)\dr_\m$ on
$Q$. It is an infinitesimal generator of a one-parameter group of local
diffeomorphisms of $Q$. Since $Z_Q\to Q$ is a trivial bundle, this
vector field gives rise to a vector field $u=u^\m\dr_\m$ on $Z_Q$, and its
jet prolongation (\ref{s52}) onto
$J^1Z_Q$ reads
\mar{s59}\beq
u= u^\m \dr_\m - z^A_\nu\dr_\m u^\nu \dr_A^\m= u^\m d_\m +[-
u^\nu z^A_\nu\dr_A - d_\m (u^\nu z^A_\nu)\dr_A^\m].
\label{s59}
\eeq
One can regard it as a generalized vector field depending on parameter
functions $u^\m(q^\nu)$. In accordance with Proposition 1, 
it seems reasonable to require that, in order to describe jets of
submanifolds of $Z$, a Lagrangian $L$ on $J^1Z_Q$ is independent on
coordinates of $Q$ and must be variationally invariant under $u$
(\ref{s59}) or, equivalently, its vertical part
\be
u_V= -
u^\nu z^A_\nu\dr_A - d_\m(u^\nu z^A_\nu)\dr_A^\m.
\ee
Then the variational derivatives of this Lagrangian obey 
irreducible Noether identities (\ref{s56}) which read
\mar{s60}\beq
\bx{z^A_\nu\cE_A=0.} \label{s60}
\eeq
These Noether identities are rather restrictive, unless other fields are
introduced.

In order to extend Lagrangian formalism on $Z_Q$ to other
fields, one can use the following two constructions:
(i) a bundle product
$Z_Q\times Z'$ of $Z_Q$ and some bundle $Z'\to Q$ bundle over
$Q$ (e.g., a tensor bundle $\op\ot^kTQ\op\ot^rT^*Q$), (ii) a bundle
$E\to Z$ and  its pull-back $E_Q$ onto $Q\times Z$ which is a composite
bundle
\mar{s61}\beq
E_Q=Q\times E\to Z_Q \to Q. \label{s61}
\eeq 

Let $E\to Z$ be provided with bundle coordinates $(z^A, s^i)$. Its
pull-back $E_Q$ onto $Z_Q$ possesses coordinates $(q^\m,z^A,s^i)$.
Accordingly, the pull-back
$Q\times J^1E$ of the first order jet manifold
$J^1E$ of
$E\to Z$ onto
$Z_Q$ is endowed with coordinates $(q^\m,z^A, s^i, s^i_A)$. It is a
subbundle $Q\times J^1E\subset J^1E_Q$ of the first order jet manifold
$J^1E_Q\to Z_Q$ of the fiber bundle $E_Q\to Z_Q$. This subbundle consists
of jets of sections of $E_Q\to Z_Q$ which are the pull-back of sections of
$E\to Z$. Given a composite fiber bundle (\ref{s61}), there is the
canonical bundle morphism
\mar{s62}\beq
\g: J^1Z_Q\op\times_{Z_Q} J^1E_Q \op\to_{Z_Q} J^1_QE_Q \label{s62}
\eeq
of the bundle product of jet manifolds $J^1Z_Q$, $J^1E_Q$ of the bundles
$Z_Q\to Q$, $E_Q\to Z_Q$ to the first order jet manifold $J^1_QE_Q$ of
the fiber bundle $E_Q\to Q$ \cite{book,sau,book00}. The jet
manifold $J^1_QE_Q$ is coordinated by $(q^\m,z^A,s^i,z^A_\m,s^i_\m)$.    
Restricted to
$Q\times J^1E\subset J^1E_Q$, the morphism (\ref{s62}) takes the
coordinate form  
\mar{s63}\beq
(z^A_\m,s_\m^i)=\g(z^A_\m,s^i_A)=(z^A_\m,s_A^iz^A_\m). \label{s63}
\eeq
Due to the morphism (\ref{s63}), any connection 
\mar{s64}\beq
\G=dz^A\ot(\dr_A + \G^i_A(z^B,s^j)\dr_i) \label{s64}
\eeq
on a fiber bundle $E\to Z$ yields the covariant derivative
\mar{s65}\beq
D_\m s^i=s^i_\m - \G^i_A(z^B,s^j)z^A_\m \label{s65}
\eeq
on the composite bundle $E_Q\to Q$.

Given a fiber bundle $Z_Q\times Z'$ or a composite fiber bundle
$E_Q$ (\ref{s61}), an extended Lagrangian is defined on the
jet manifolds $J^1Z_Q\times J^1Z'$ or $J^1_QE_Q$, respectively. For
instance, any horizontal $n$-form
\be
\frac1{n!}\f_{A_1\ldots A_n}dz^{A_1}\w\cdots\w dz^{A_n}
\ee
on $E\to Z$ yields a horizontal density
\be
\frac1{n!}\f_{A_1\ldots A_n}z^{A_1}_{\m_1}\cdots
z^{A_n}_{\m_n}dq^{\m_1}\w\cdots\w dq^{\m_n}
\ee 
on $J^1_QE_Q\to Q$ which may contribute to a Lagrangian (see, 
e.g., a relativistic particle in
the presence of an electromagnetic field).

\section{Hamiltonian dynamics of submanifolds} 

Here, we follow
polysymplectic Hamiltonian formalism which aims to describe
field systems with nonregular Lagrangians
\cite{book,jpa99,sard94,book95}.  Lagrangian and polysymplectic
Hamiltonian formalisms are equivalent in the case of hyperregular
Lagrangians, but a nonregular Lagrangian admits different associated
Hamiltonians, if any. At the same time, there is a comprehensive relation
between these formalisms in the case of almost-regular Lagrangians. 

Given a fiber bundle $Z_Q$ (\ref{s12}) and its vertical cotangent bundle
$V^*Z_Q$, let us consider the fiber bundle
\mar{s70}\beq
\Pi=V^*Z_Q\w (\op\w^{n-1} T^*Q). \label{s70}
\eeq
It plays the role of a momentum phase space
of covariant Hamiltonian field theory.  Given coordinates
$(q^\m,z^A)$ on
$Z_Q$, this fiber bundle is coordinated by $(q^\m, z^A, p^\m_A)$, where
$p^\m_A$ are treated as coordinates of momenta. It is provided with the
canonical polysymplectic form
\mar{s73}\beq
\Om_\Pi= dp_A^\m\w dz^A\ot \dr_\m. \label{s73}
\eeq

Every Lagrangian $L$ on the jet manifold
$J^1Z_Q$ yields the Legendre map 
\mar{s71}\beq
\wh L: J^1Z_Q\op\to_{Z_Q} \Pi, \qquad p^\m_A\circ\wh L =\dr^\m_A\cL,
\label{s71}
\eeq 
whose range $N_L=\wh L(J^1Z_Q)$ is called the Lagrangian constraint
space. A Lagrangian $L$ is called hyperregular (resp. regular), if the
Legendre map (\ref{s71}) is a diffeomorphism (resp. local diffeomorphism,
i.e., of maximal rank). A Lagrangian
$L$ is said to be almost-regular if the Lagrangian constraint space is a
closed imbedded subbundle
$i_N:N_L\to\Pi$ of the Legendre bundle $\Pi\to Z_Q$ and the surjection
$\wh L:J^1Z_Q\to N_L$ is a fibered manifold possessing connected fibers.

A multisymplectic momentum phase space is the homogeneous Legendre
bundle
\be
\ol \Pi=T^*Z_Q\w (\op\w^{n-1} T^*Q),
\ee
coordinated by $(q^\m, z^A, p^\m_A, p)$. It is endowed with the canonical
multisymplectic form
\be
\Xi=p\om + p^\m_A dz^A\w\om_\m.
\ee
There is a trivial one-dimensional bundle $\ol\Pi\to \Pi$. 
Then a Hamiltonian $\cH$ on the momentum phase space $\Pi$ (\ref{s70}) is
defined as a section
$p=-\cH$ of this fiber bundle. The pull-back of the multisymplectic form
$\Xi$ onto
$\Pi$ by a Hamiltonian $\cH$ is a Hamiltonian form
\mar{s75}\beq
H= p^\m_Adz^A\w\om_\m -\cH\om \label{s75}
\eeq 
on $\Pi$. The corresponding Hamilton equations with respect to the
polysymplectic form $\Om_\Pi$ (\ref{s73}) read
\be
z^A_\m-\dr^A_\m\cH=0, \qquad -p^\m_{\m A}- \dr_A\cH=0.
\ee
A key point is that these Hamilton equations coincide with
the Euler--Lagrange equations of the first order Lagrangian
\mar{s77}\beq
L_H= (p^\m_Az^A_\m - \cH) \label{s77}
\eeq
on the jet manifold $J^1\Pi$ of $\Pi\to Q$. Indeed, its variational
derivatives are
\mar{s101}\beq
\cE^A_\m=z^A_\m-\dr^A_\m\cH, \qquad \cE_A=-p^\m_{\m A}- \dr_A\cH.
\label{s101}
\eeq

Any Hamiltonian form $H$ (\ref{s75}) on $\Pi$ yields the Hamiltonian map
\mar{s80}\beq
\wh H: \Pi\op\to_{Z_Q} J^1Z_Q, \qquad  z^A_\m\circ \wh H= \dr^A_\m\cH.
\label{s80}
\eeq
A Hamiltonian $\cH$ on $\Pi$ is said to be associated to a Lagrangian $L$
on $J^1Z_Q$ if it satisfies the relations
\mar{s81,2}\ben
&& p^\m_A=\dr^\m_A\cL(q^\nu, z^B, z^B_\la=\dr^B_\la\cH), \label{s81}\\
&& p^\m_A\dr_\m^A\cH -\cH=\cL(q^\nu, z^B, z^B_\la=\dr^B_\la\cH).
\label{s82}
\een
If an associated Hamiltonian $\cH$ exists, the Lagrangian constraint space
$N_L$ is given by the coordinate equalities (\ref{s81}).
The relation between Lagrangian and polysymplectic Hamiltonian
formalisms is based on the following facts. 

(i) Let a Lagrangian $L$ be
almost regular, and let us assume that it admits an associated
Hamiltonian $\cH$, which however need not be unique, unless $L$ is
hyperregular. In
this case, the Poincar\'e--Cartan form
$H_L$ (\ref{s43}) is the pull-back $H_L=\wh L^*H$ of the Hamiltonian form
$H$ (\ref{s75}) for any associated Hamiltonian $\cH$. Note that a local
associated Hamiltonian always exists. The Poincar\'e--Cartan form $H$ is
a Lepagean equivalent both of the original Lagrangian $L$ on $J^1Z_Q$ and
the Lagrangian
\be
\ol L=(\cL + (\wh z^A_\m - z^A_\m)\dr^\m_A\cL)\om
\ee
on the repeated jet manifold $J^1J^1Z_Q$. Its Euler--Lagrange equations
are the Cartan equations for $L$. Any solution of the Euler--Lagrange
equations (\ref{s90}) for $L$ is also a solution of the Cartan equations.
Furthermore, Euler--Lagrange equations and the Cartan equations are
equivalent in the case of a regular Lagrangian. 

(ii) If a Lagrangian $L$ is almost regular, all
associated Hamiltonian forms $H$ coincide with each other on the
Lagrangian constraint space $N_L$, and define the constrained Lagrangian
$L_N= h_0(i_N^*H)$ on the jet manifold $J^1N_L$ of the fiber bundle
$N_L\to Q$. The Euler--Lagrange equations for this Lagrangian are called
the constrained Hamilton equations. In fact, the Lagrangian $L_H$
(\ref{s77}) is defined on the bundle product 
\mar{s100}\beq
\Pi\op\times_{Z_Q} J^1Z_Q, \label{s100}
\eeq
 and the constrained Lagrangian $L_N$ is the restriction of
$L_H$ to $N_L\times J^1Z_Q$.

As a result, one can show that a section $\ol
S$ of the jet bundle $J^1Z_Q\to Q$ is a solution of the Cartan equations
for $L$ iff $\wh L\circ \ol S$ is a solution of the constrained Hamilton
equations. In particular, any solution $r$ of the constrained Hamilton
equations provides the solution $\ol S=\wh H\circ r$ of the Cartan
equations. 

Turn now to symmetries of a Lagrangian $L_H$ (\ref{s77}). Any vector
field $u$ on $Z_Q$ gives rise to the vector field
\be
u_\Pi= u^\m\dr_\m + u^A\dr_A +(-\dr_A u^B p^\m_B -\dr_\la u^\la p^\m_A
+\dr_\la u^\m p^\la_A)\dr^A_\m
\ee
onto the Legendre bundle $\Pi$. Then we obtain its prolongation 
\be
u_\Pi= u^\m\dr_\m + u^A\dr_A + (d_\m u^A -z^A_\nu d_\m
u^\nu)\dr_A^\m + (-\dr_A u^B p^\m_B -\dr_\la u^\la p^\m_A +\dr_\la u^\m
p^\la_A)\dr^A_\m
\ee
onto the product (\ref{s100}). It is a variational symmetry of the
Lagrangian $L_H$ if the Lie derivative $\bL_{u_\Pi}L_H$ is $d_H$-exact.

For instance, let $u=u^\m\dr_\m$ be an arbitrary vector field on $Q$. 
Since $Z_Q\to Q$ is a trivial bundle, this
vector field gives rise to a vector field $u=u^\m\dr_\m$ on $Z_Q$ whose
lift onto the Legendre bundle $\Pi$ is
\be
u_\Pi= u^\m\dr_\m + (-\dr_\la u^\la p^\m_A +\dr_\la u^\m
p^\la_A)\dr^A_\m.
\ee
Then we obtain its prolongation 
\be
u_\Pi=u^\m\dr_\m - z^A_\nu\dr_\m u^\nu\dr^\m_A + (-\dr_\la u^\la p^\m_A
+\dr_\la u^\m p^\la_A)\dr^A_\m
\ee
onto the product (\ref{s100}), and take its vertical part
\mar{s105}\beq
u_V= -u^\nu z^A_\nu\dr_A - d_\m(u^\nu z^A_\nu)\dr_A^\m + (-\dr_\la u^\la
p^\m_A +\dr_\la u^\m p^\la_A -u^\nu p_{\nu A}^\m)\dr^A_\m. \label{s105}
\eeq
Let us regard it as a generalized vector field dependent on parameter
functions $u^\m(q)$. In accordance with Proposition 1, let us require
that a Lagrangian $L_H$ is independent on coordinates on $Q$ and
possesses the gauge symmetry $u_V$ (\ref{s105}). Then its 
variational derivatives (\ref{s101}) of $L_H$ obey
the Noether identities
\be
z^A_\nu \cE_A+ p^\m_{\m A} \cE^A_\nu + p^\m_A(d_\m\cE_\nu^A -
d_\nu\cE^A_\m) =0, 
\ee
which reduce to rather restrictive conditions
\mar{s110}\beq
\bx{\dl^\m_\nu \cH =(n-1) p^\m_A\dr^A_\nu\cH}
\label{s110}
\eeq
which a Hamiltonian $\cH$ must satisfy. For instance, $\cH=0$ if $n=1$.
In this case, momenta are scalars relative to transformations of $q$
and, therefore, no function of them is a density  with respect to these
transformations.

\section{Example. $n=1,2$}

Given an $m$-dimensional manifold $Z$ coordinated by $(z^A)$, let us
consider the jet manifold $J^1_1Z$ of its one-dimensional submanifolds.
Let us provide $Z=Z^0_1$ with coordinates $(x^0=z^0, y^i=z^i)$
(\ref{5.8}). Then the jet manifold $J^1_1Z$ is endowed with coordinates 
$(z^0,z^i,z^i_0)$ possessing transition functions (\ref{5.26}),
(\ref{s2}) which read
\mar{s120}\beq
z'^0=z'^0(z^0,z^k), \quad z'^0=z'^0(z^0,z^k), \quad z'^i_0=
(\frac{\dr z'^i}{\dr z^j} z^j_0 + \frac{\dr z'^i}{\dr z^0} )
(\frac{\dr z'^0}{\dr z^j} z^j_0 + \frac{\dr z'^0}{\dr z^0} )^{-1}.
\label{s120}
\eeq 
A glance at the transformation law (\ref{s120}) shows that $J^1_1Z\to Z$
is a fiber bundle in projective spaces.

For instance, put $Z=\Bbb R^4$ whose Cartesian coordinates are subject
to the Lorentz transformations
\mar{s122}\beq
z'^0= z^0{\rm ch}\al - z^1{\rm sh}\al, \qquad z'^'= -z^0{\rm sh}\al +
z^1{\rm ch}\al, \qquad z'^{2,3} = z^{2,3}. \label{s122}
\eeq
Then $z'^i$ (\ref{s120}) are exactly the Lorentz transformations
\be
z'^1_0=\frac{ z^1_0{\rm ch}\al -{\rm sh}\al}{ -  z^1_0{\rm
sh}\al+ {\rm ch}\al} \qquad 
z'^{2,3}_0=\frac{z^{2,3}_0}{ - z^1_0{\rm
sh}\al + {\rm ch}\al}
\ee
of three-velocities in relativistic mechanics \cite{sard98,book98}.

Let us consider a one-dimensional manifold $Q=\Bbb R$ 
and the product $Z_Q=\Bbb R\times Z$. Let $\Bbb R$ be provided with a
Cartesian coordinate $\tau$ possessing transition function $\tau'=\tau +
{\rm const}$, unless otherwise stated. 
Then the jet manifold
$J^1Z_Q$ of the fiber bundle  $\Bbb R\times Z\to Z$ is endowed with the
coordinates $(\tau, z^0, z^i, z^0_\tau, z^i_\tau)$ with the transition
functions
\mar{s123}\beq
z'^0_\tau=\frac{\dr z'^0}{\dr z^k} z^k_\tau + \frac{\dr z'^0}{\dr z^0}
z^0_\tau, \qquad 
z'^i_\tau=\frac{\dr z'^i}{\dr z^k} z^k_\tau + \frac{\dr z'^i}{\dr z^0}
z^0_\tau. \label{s123}
\eeq
A glance at this transformation law shows that, unless
nonadditive transformations of
$\tau$ are considered, there is
an isomorphism 
\mar{s121}\beq
J^1_1Z_Q =VZ_Q=\Bbb R\times TZ  \label{s121}
\eeq
of the jet manifold $J^1_1Z_Q$ to the vertical tangent bundle $VZ_Q$ of
$Z_Q\to \Bbb R$ which, in turn, is a product of
$\Bbb R$ and the tangent bundle $TZ$ of $Z$.

Returning to the example of $Z=\Bbb R^4$ and Lorentz transformations
(\ref{s122}), one easily observed that transformations (\ref{s123}) are
transformations of four-velocities in relativistic mechanics where $\tau$
is a proper time.

Let us consider coordinate charts $(U';\tau, z^0,z^i,z^i_0)$ and
$(U'';\tau, z^0,z^i,z^0_\tau, z^i_\tau)$ of the manifolds $\Bbb R\times
J^1_1Z$ and 
$J^1Z_Q$ over the same coordinate chart $(U;\tau, z^0,z^i)$ of $Z_Q$. 
Then one can associate to each element $(\tau, z^0,z^i,z^i_0)$ of
$U'\subset \Bbb R\times
J^1_1Z$ the elements of $U''\subset J^1Z_Q$ which obey the relations
\mar{s125}\beq
z^i_0 z^0_\tau= z^i_\tau \label{s125}
\eeq
and, in particular, the relations
\mar{s126}\beq
z^i_0= \frac{z^i_\tau}{z^0_\tau}, \qquad z^0_\tau\neq 0. \label{s126}
\eeq 
Given a point $(\tau,z)\in \Bbb R\times
Z$, the relations (\ref{s125}) -- (\ref{s126}) are exactly the
correspondence between elements of a one-dimensional vector subspace of
the tangent space $T_zZ$ and the corresponding element of the projective
space of these subspaces. 

In the above mentioned example of relativistic
mechanics, the relations (\ref{s125}) -- (\ref{s126}) are familiar
equalities between three- and four-velocities. It should be emphasized
that, in relativistic mechanics, one avoids the ambiguity between
three- and four-velocities by means of the nonholonomic constraint
\mar{s131}\beq
(z^0_\tau)^2- \op\sum_i(z^i_\tau)^2 =1. \label{s131}
\eeq

In a general setting, Lagrangian formalism on the jet manifold $J^1Z_Q$
can be developed if a Lagrangian $L$ is independent of $\tau$, and it is
variationally invariant under transformations reparametrizations
$\tau'(\tau)$, i.e.,  its Euler--Lagrange operator obeys the Noether
identity 
\mar{s130}\beq
z^A_\tau\cE_A=0.  \label{s130}
\eeq

For instance, let $Z$ be a locally affine manifold, i.e., a toroidal
cylinder $\Bbb R^{m-k}\times T^k$. Its tangent bundle can be provided
with a constant nondegenerate fiber metric $\eta_{AB}$. Then 
\mar{s132}\beq
L=(\eta_{AB}z^A_\tau z^B_\tau)^{1/2}d\tau \label{s132}
\eeq
is a
Lagrangian on $J^1Z_Q$. It is easily justified that this Lagrangian
satisfies the Noether identity (\ref{s130}). Furthermore, given a one-form
$\cA_Bdz^B$ on $Z$, one can consider the Lagrangian
\label{s133}\beq
L'=[(\eta_{AB}z^A_\tau z^B_\tau)^{1/2} - \cA_Bz^B_\tau]d\tau, \label{s133}
\eeq
which also obeys the Noether identity (\ref{s130}).
In relativistic mechanics, the Euler--Lagrange equations of the
Lagrangians $L$ (\ref{s132}) and $L'$ (\ref{s133}) restricted to the
constraint space (\ref{s131}) restart the familiar equations of motion of
a free relativistic particle particle and a relativistic particle in the
presence of an electromagnetic field $\cA$.

As was mentioned above, no Hamiltonian obeys the Noether identities
(\ref{s110}) if $n=1$. However, Hamiltonian relativistic
mechanics can be developed in the framework of Hamiltonian theory of
mechanical systems with nonholonomic constraints
\cite{book98,sard99,sard03}. A key is that the constraint condition
(\ref{s131}) is not preserved under transformations of $\tau$, and a
Hamiltonian of a mechanical system with this constraint need not
satisfy the  Noether identities (\ref{s110}).

In comparison with the case of one-dimensional submanifolds, a
description of the Lagrangian and Hamiltonian dynamics of two-dimensional
submanifolds follows general theory of $n$-dimensional submanifolds. This
is the case of classical string theory
\cite{scherk,hatf,polch}.  

For instance, let $Z$ be again an $m$-dimensional locally affine manifold,
i.e., a toroidal cylinder $\Bbb R^{m-k}\times T^k$,
and let $Q$ be a
two-dimensional manifold. As was mentioned above, the tangent bundle 
of $Z$ can be provided with a
constant nondegenerate fiber metric
$\eta_{AB}$. Let us consider the $2\times 2$ matrix with the
entries
\be
h_{\m\nu}=\eta_{AB} z^A_\m z^B_\nu.
\ee
Then its determinant provides a 
Lagrangian
\mar{s140}\beq
L=(\det h)^{1/2} d^2q =([\eta_{AB} z^A_1 z^B_1] [\eta_{AB} z^A_2
z^B_2]- [\eta_{AB} z^A_1 z^B_2]^2 )^{1/2} d^2q  \label{s140}
\eeq
on the jet manifold  $J^1Z_Q$ (\ref{s30}). This is the well known
Nambu--Goto Lagrangian of string theory. It satisfies the Noether
identities (\ref{s60}). Let 
\be
F=\frac12 F_{AB} dz^A\w dz^B
\ee
be a two-form on a manifold $Z$.
Then 
\be
F=\frac12 F_{AB} z^A_\m z^B_\nu dq^\m\w dq^\nu
\ee
is a horizontal density on $J^1Z_Q$ which can be treated as an
interaction term of submanifolds and an external classical field $F$ in
a Lagrangian. 

Turn now to Hamiltonian theory of two-dimensional submanifolds on the
momentum phase space $\Pi$ (\ref{s70}). In this case, the Noether
identities (\ref{s110}) take the form
\mar{s145}\beq
\dl^\m_\nu\cH=p^\m_A\dr^A_\nu\cH. \label{s145}
\eeq
For instance, let $Z$ be the above mentioned toroidal cylinder whose
cotangent bundle is provided with a constant nondegenerate fiber metric
$\eta_{AB}$.  Let us consider the $2\times 2$ matrix with the
entries
\be
H^{\m\nu}=\eta^{AB} p_A^\m p_B^\nu.
\ee
Then its determinant provides a Hamiltonian 
\be
\cH=(\det H)^{1/2} d^2q =([\eta^{AB} p_A^1 p_B^1] [\eta^{AB} p_A^2
p_B^2]- [\eta^{AB} p_A^1 p_B^2]^2 )^{1/2} d^2q
\ee
on the momentum phase space $\Pi$ which satisfies the Noether identities
(\ref{s145}. This Hamiltonian is associated to the Lagrangian
(\ref{s140}).


\begin{thebibliography}{ddd}

\bibitem{tak2} F.Takens, A global version of the inverse problem of
the calculus of variations, {\it J. Diff. Geom.} {\bf 14} (1979)
543.

\bibitem{ander} I.Anderson, Introduction to the variational bicomplex,
{\it Contemp. Math.}, {\bf 132} (1992) 51. 

\bibitem{jmp01} G.Giachetta, L.Mangiarotti and G.Sardanashvily, Cohomology
of the infinite-order jet space and the inverse problem, {\it J. Math. Phys.}
{\bf 42} (2001) 4272.

\bibitem{barn} G.Barnish, F.Brandt and M.Henneaux, Local BRST cohomology
in gauge theories, {\it Phys. Rep.} {\bf 338} (2000) 439.

\bibitem{cmp04} G.Giachetta, L.Mangiarotti and G.Sardanashvily,
Lagrangian supersymmetries depending on derivatives. Global
analysis and cohomology. {\it Commun. Math. Phys.} {\bf 259} (2005) 103;
{\it E-print arXiv}: math.AG/0305303.


\bibitem{jmp05} D.Bashkirov, G.Giachetta, L.Mangiarotti and
G.Sardanashvily, Noether's second theorem for BRST symmetries.
{\it J. Math. Phys.} {\bf 46} (2005) 053517; {\it E-print arXiv:}
math-ph/0412034.

\bibitem{book} G.Giachetta, L.Mangiarotti and G.Sardanashvily, {\it New
Lagrangian and Hamiltonian Methods in Field Theory} (World Scientific,
Singapore, 1997).


\bibitem{jpa99} G.Giachetta, L.Mangiarotti and G.Sardanashvily, Covariant
Hamilton equations for field theory, {\it J. Phys. A} {\bf 32} (1999)
6629; {\it E-print arXiv}: hep-th/9904062.

\bibitem{krup} O.Krupkova, Hamiltonian field theory, {\it J. Geom. Phys.}
{\bf 43} (2002) 93.

\bibitem{ech} A.Echeverr\'{\i}a Enr\'{\i}quez,G.L\'opez, J.Marin-Solano,
M.Mu\~noz Lecanda and N.Rom\'an Roy, Lagrangian-Hamiltonian unified
formalism for field theories, {\it J. Math. Phys.} {\bf 45} (2004) 360.

\bibitem{leon} M de Leon, D.Mart\'{\i}n de Diego and
A.Santamar\'{\i}a-Merini, Symmetries in classical field theory, {\it Int.
J. Geom. Methods. Mod. Phys.} {\bf 1} (2004) 651.


\bibitem{kras} I.Krasil'shchik, V.Lychagin and A.Vinogradov, {\it
Geometry of Jet Spaces and Nonlinear Partial Differential Equations}
(Gordon and breach, Glasgow, 1985)

\bibitem{modu} M.Modugno and A.Vinogradov, Some variations on the
notion of connections, {\it Ann. Matem. Pura ed Appl.} {\bf CLXVII}
(1994) 33.


\bibitem{sard98} G.Sardanashvily, Hamiltonian time-dependent mechanics,
{\it J. Math. Phys.} {\bf 39} (1998) 2714.

\bibitem{book98} L.Mangiarotti and G.Sardanashvily, {\it Gauge
Mechanics} (World Scientific, Singapore, 1998).


\bibitem{ste} N.Steenrod, {\it The Topology of Fibre Bundles} (Princeton Univ.
Press, Princeton, 1972).


\bibitem{jpa05} D.Bashkirov, G.Giachetta,  L.Mangiarotti and
G.Sardanashvily, Noether's second theorem in a general setting.
Reducible gauge theories, {\i J. Phys. A} {\bf 38} (2005) 5329; 
{\it E-print arXiv:} math.DG/0411070

\bibitem{jmp05a} D.Bashkirov, G.Giachetta, L.Mangiarotti and
G.Sardanashvily, The antifield Koszul--Tate complex of reducible
Noether identities. {\it J. Math. Phys.} {\bf 46} (2005) 103513;
{\it E-print arXiv:} math-ph/0506034.

\bibitem{sard94} G.Sardanashvily, Constraint field systems in
multimomentum canonical variables, {\it J. Math. Phys.} {\bf  35} (1994)
6584. 

\bibitem{book95} G.Sardanashvily, {\it Generalized Hamiltonian Formalism
for Field theory} (World Scientific, Singapore, 1995).

\bibitem{sau} D.Saunders, {\it The Geometry of Jet Bundles}
(Cambr. Univ. Press, Cambridge, 1989).


\bibitem{book00} L.Mangiarotti and G.Sardanashvily, {\it Connections in
Classical and Quantum Field Theory} (World Scientific, Singapore, 2000).


\bibitem{sard99} G.Giachetta, L.Mangiarotti and
G.Sardanashvily, Nonholonomic constraints in time-dependent mechanics
{\it J. Math. Phys.} {\bf 40} (1999) 1375; {\it  E-print arXiv}:
math-ph/9807014.

\bibitem{sard03} G.Sardanashvily, Geometric quantization of relativistic
Hamiltonian mechanics, {\it Int. J. Theor. Phys.} {\bf 42} (2003) 697; 
{\it  E-print arXiv}:
gr-qc/0208073.


\bibitem{scherk} J.Scherk, An introduction to the theory of dual models
and strings, {\it Rev. Mod. Phys.} {\bf 47} (1975) 123.

\bibitem{hatf} B.Hatfield, {\it Quantum Field Theory of Point Particles
and Strings} (Addison--Willey Publ., Redwood City, CA, 1992).

\bibitem{polch} J.Polchinski, {\it String Theory} (Cambr. Univ. Press,
Cambridge, 1998).



\end{thebibliography}
\end{document}